\documentclass[conference]{IEEEtran}
\IEEEoverridecommandlockouts
\usepackage{cite}
\usepackage{amsmath,amssymb,amsfonts}
\usepackage{algorithmic}
\usepackage{graphicx}
\usepackage{textcomp}
\usepackage{xcolor}
\usepackage{textcomp}
\usepackage{multirow}
\usepackage{epsfig}
\usepackage{subfigure}
\usepackage{xcolor}
\usepackage{multirow}
\usepackage{amsmath,amssymb,amsfonts}
\usepackage{xcolor,color,colortbl}
\usepackage{amsmath,amssymb,amsfonts}
\usepackage{algorithm}
\usepackage{enumitem}

\usepackage{subfigure}
\usepackage{stfloats}
\usepackage{array}
\usepackage{epsfig}
\usepackage{amssymb}
\usepackage{changes}
\usepackage{textcomp}
\usepackage{siunitx}
\newcommand{\bigO}{\mathcal{O}}
\usepackage{algorithmic}
\newcommand\myeq{\mathrel{\stackrel{\makebox[0pt]{\mbox{\normalfont\tiny (a)}}}{=}}}

\usepackage{balance}

\def\BibTeX{{\rm B\kern-.05em{\sc i\kern-.025em b}\kern-.08em
 T\kern-.1667em\lower.7ex\hbox{E}\kern-.125emX}}
\begin{document}

\title{Deep Learning-Based Optimal RIS Interaction Exploiting Previously Sampled Channel Correlations}

\author{\IEEEauthorblockN{Mehmet Ali Ayg\"{u}l\IEEEauthorrefmark{1},
 Mahmoud~Nazzal\IEEEauthorrefmark{1},~H\"{u}seyin~Arslan\IEEEauthorrefmark{1}\IEEEauthorrefmark{2}}
\IEEEauthorblockA{\IEEEauthorrefmark{1}Department of Electrical and Electronics Engineering, Istanbul Medipol University, Istanbul, 34810, Turkey}
\IEEEauthorblockA{\IEEEauthorrefmark{2}Department of Electrical Engineering, University of South Florida, Tampa, FL, 33620, USA}
\\
Emails: \{mehmetali.aygul, mahmoud.nazzal\}@ieee.org, arslan@usf.edu

\thanks{This work has been submitted to the IEEE for possible publication. Copyright may be transferred without notice, after which this version may no longer be accessible.}}

\maketitle

\begin{abstract}
The reconfigurable intelligent surface (RIS) technology has attracted interest due to its promising coverage and spectral efficiency features. However, some challenges need to be addressed to realize this technology in practice. One of the main challenges is the configuration of reflecting coefficients without the need for beam training overhead or massive channel estimation. Earlier works used estimated channel information with deep learning algorithms to design RIS reflection matrices. Although these works can reduce the beam training overhead, still they overlook existing correlations in the previously sampled channels. In this paper, different from existing works, we propose to exploit the correlation in the previously sampled channels to estimate RIS interaction more reliably. We use a deep multi-layer perceptron for this purpose. Simulation results reveal performance improvements achieved by the proposed algorithm.
\end{abstract}


\section{Introduction}
\label{Section1}

\par The massive multiple-input multiple-output (mMIMO) technology is a key enabler for communications in the fifth-generation and beyond that ensures high spectral efficiency exploiting the high spatial multiplexing gains \cite{5Genvision, pan2020intelligent}. This technology is especially useful when it operates at millimeter-wave frequencies. Reconfigurable intelligent surfaces (RISs) \cite{basar2019wireless,gong2020towards} have been proposed as a means of improving the coverage and efficiency of mMIMO systems.

\par An RIS is structured to have a massive number of passive reflecting elements. Each element reflects an incident wave in a controllable direction and phase. Similar to beamforming at the base station (BS), RIS operation in optimizing the communicating requires precisely knowing the wireless channel. Therefore, channel estimation is an integral part of RIS interaction \cite{el2020reconfigurable}. An outstanding challenge facing channel estimation in this context is the huge number of reflecting elements. This leads to prohibitively high channel estimation and hardware complexity burdens if traditional methods are to be employed \cite{zheng2019intelligent} in this setting.

\par Prior art on RIS interaction addresses the problems of channel estimation and beamforming design \cite{taha2019enabling,huang2019indoor,elbir2020deep,gao2020unsupervised,jensen2020optimal,taha2020deep}. Along this line, the overhead of channel and beam training is addressed in \cite{taha2019enabling} through the use of compressive sensing and deep learning (DL). Subsequently, an approach for efficient RIS configuration is proposed in \cite{huang2019indoor}. The promising gains of these solutions motivated more research in these directions \cite{elbir2020survey}. More recently, supervised DL is used to exploit the mapping between pilots to improve channel estimation \cite{elbir2020deep}, and unsupervised DL to reflect beamforming for RIS-assisted systems \cite{gao2020unsupervised}. Besides, a minimum variance unbiased estimator is used to leverage channel estimation in \cite{jensen2020optimal}. Moreover, deep reinforcement learning is used to estimate the best reflection coefficients of the RIS surface by adjusting its reflection matrix \cite{taha2020deep}.

\par Despite the successes of the aforementioned approaches, they overlook existing correlations in previous channel information. Inherently, a strong correlation exists between previously sampled channels and the ones being estimated. The exploitation of this correlation naturally promises to improve the performance of RIS interaction.

\par This paper proposes an algorithm for exploiting previous channel information for improving the quality of optimal RIS interaction. We design a deep multi-layer perceptron (MLP) for this purpose. Extensive experimental results show a performance improvement achieved by the proposed algorithm. These results are validated over the DeepMIMO dataset \cite{alkhateeb2019deepmimo}.

\par \textit{Organization:} The rest of the paper is organized as follows. The system model and preliminaries are presented in Section~\ref{Section2}. Section~\ref{Section3} details the proposed algorithm for optimum phase interaction. Simulation results are presented in Section~\ref{Section4}, and the paper is concluded in Section~\ref{Section5}.

\par \textit{Notation:} Plain-faced, bold-faced lower-case, bold-faced upper-case, and calligraphic font letters represent scalars, vectors, matrices, and sets of vectors, respectively. A determinant of a matrix $\boldsymbol{X}$ is denoted by ${|\boldsymbol{X}|}$. $\mathbb{E}[.]$ represents statistical expectation. The superscript $T$ denotes algebraic transpose. $\boldsymbol{A} \odot \boldsymbol{B}$ is the Hadamard product of $\boldsymbol{A}$ and $\boldsymbol{B}$. $\mathcal{N}(m,R)$ is a complex Gaussian random vector with mean $m$ and covariance $R$.

\section{System Model and Preliminaries} 
\label{Section2}

\subsection{System Model} 

\par This work assumes a $K$-subcarrier OFDM communication system. Communication takes place between single-antenna transceiver ends. An $M$-element RIS is considered to support this communication. Similar to the case in \cite{taha2019enabling}, we assume that there is no direct link for communication to simplify the setting. Thus, the signal observed at the receiver end can be written as follows 
\begin{IEEEeqnarray}{rCl}
y_k & = & \boldsymbol{h}_{R,k}^T\boldsymbol{\Psi}\boldsymbol{h}_{T,k}\boldsymbol{s}_k+n_k, \label{eq1}
\\
& \myeq & (\boldsymbol{h}_{R,k}\odot \boldsymbol{h}_{T,k})^T\boldsymbol{\psi} s_k + n_k, \label{eq2}
\end{IEEEeqnarray}
\noindent where $s_k$ is the transmit signal governed by $\mathbb{E}[|s_k|^2]=\frac{P_T}{K}$, and $P_T$ is the total transmit power. Also, $\boldsymbol{\Psi}$ represents the RIS interaction diagonal matrix. $n_k\sim \mathcal{N}_\mathbb{C}(0,\sigma_{n}^{2})$ is the received noise, and $\boldsymbol{h}_{T,k}$, $\boldsymbol{h}_{R,k}\in \mathbb{C}^{M\times 1}$ is the channel from the transceiver to the RIS at the 
$k^{th}$ subcarrier.

\par
\begin{figure}[t!] 
\setlength\abovecaptionskip{-0.1\baselineskip}
\centering\resizebox{0.92\columnwidth}{!}{
\includegraphics{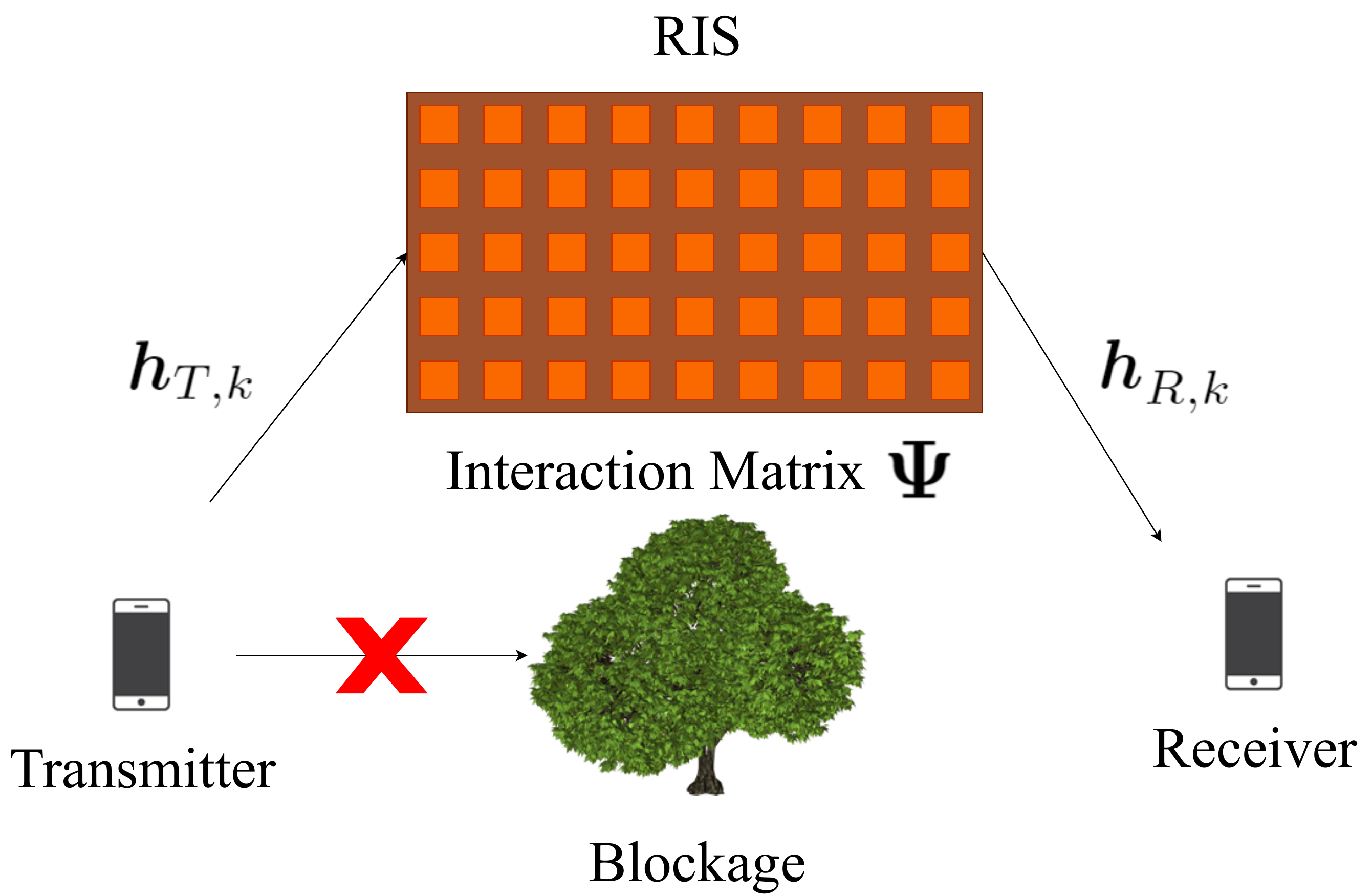}}
\caption{The system model of an RIS-assisted transceiver system.}
\label{basic}
\end{figure}

 The main system model components are shown by Fig.~\ref{basic}. It is noted that phase shifters are assumed in the RIS architecture. Thus, an interaction vector is modelled as $[\boldsymbol{\psi}]_m=e^{j\phi_m}$ and an interaction vector is chosen from a foreknown set of interaction vector codebook $\mathcal{P}$. We adopt an assumption made in \cite{taha2019enabling} that a few active reflecting elements are randomly placed between the passive ones on the RIS. Thus, the sampled channel vector from the transceiver to the RIS active elements, $\overline{\rm \boldsymbol{h}}_{T,k}$, $\overline{\rm \boldsymbol{h}}_{R,k}\in \mathbb{C}^{\overline{\rm M}}\times 1$, can be expressed as $\overline{\rm \boldsymbol{h}}_{T,k}=\boldsymbol{G}_{RIS}\boldsymbol{h}_{T,k}$ and $\overline{\rm \boldsymbol{h}}_{R,k}=\boldsymbol{G}_{RIS}\boldsymbol{h}_{R,k}$, where $\boldsymbol{G}_{RIS}$ is an $\overline{\rm M}\times M$ selection matrix whose elements are tuned according to the RIS elements. Therefore, the RIS channel vector is written as $\overline{\rm \boldsymbol{h}}_s=\overline{\rm \boldsymbol{h}}_{T,k}\odot \overline{\rm \boldsymbol{h}}_{R,k}$.

\subsection{Channel Model} 

\par We model transmitter-RIS ($\boldsymbol{h}_{T,k}$) and RIS-receiver ($\boldsymbol{h}_{R,k}$) channels according to a wideband geometric channel model used in \cite{alkhateeb2018deep}. This model assumes $L$ scattering clusters where each cluster corresponds to a specific propagation path (ray). Each ray is described by several parameters; its azimuth/elevation angles of arrival, $\theta_l$, $\phi_l \in [0, 2\pi)$, complex coefficient $\alpha_l \in \mathbb{C}$, and time delay $\tau_l \in \mathbb{E}$. The transmitter-RIS path loss is denoted by $\rho_T$. The pulse-shaping function, with $T_s$-spaced signaling, is denoted by $p(\tau)$ at time instant $\tau$. Eventually, one can express the channel in the frequency domain as follows
\begin{equation} 
\boldsymbol{h}_{T,k}=\sqrt{\frac{M}{\rho T}}\sum^{D-1}_{d=0}\sum^{L}_{l=1}\alpha_l\boldsymbol{a}(\theta_l,\phi_l)p(dT_s-\tau_l)e^{-j\frac{2\pi k}{K}d},\label{eq3} 
\end{equation}
\noindent where $\boldsymbol{a}(\theta_l,\phi_l)\in \mathbb{C}^{M \times 1}$ is the RIS array response vector. A block-fading channel model is assumed in this work, i.e., $\boldsymbol{h}_{T,k}$ and $\boldsymbol{h}_{R,k}$ stay unchanged over the channel coherence time.

\subsection{Deep Learning}

\par DL models use two or more hidden layers to have better data representation. The usage of multiple hidden layers allows magnifying intrinsic distinctive data features while suppressing irrelevant information at each layer. Thus, raw data can be used without the requirement of sophisticated feature engineering/crafting.

\par MLP is a suitable DL model to handle grid-like data in both one or two dimensions. In this paper, two-dimensional data is aimed to be represented, so MLP is a suitable choice for this purpose. MLP with a deep architecture includes an input layer, at least two hidden layers, and an output layer. It uses backpropagation for training which is a supervised learning technique \cite{rosenblatt1961principles}. Besides, it can distinguish data of nonlinear separability \cite{cybenko1989approximation}.

\begin{figure*}[t!] 
\centering
\centering\resizebox{1.95\columnwidth}{!}{
\includegraphics{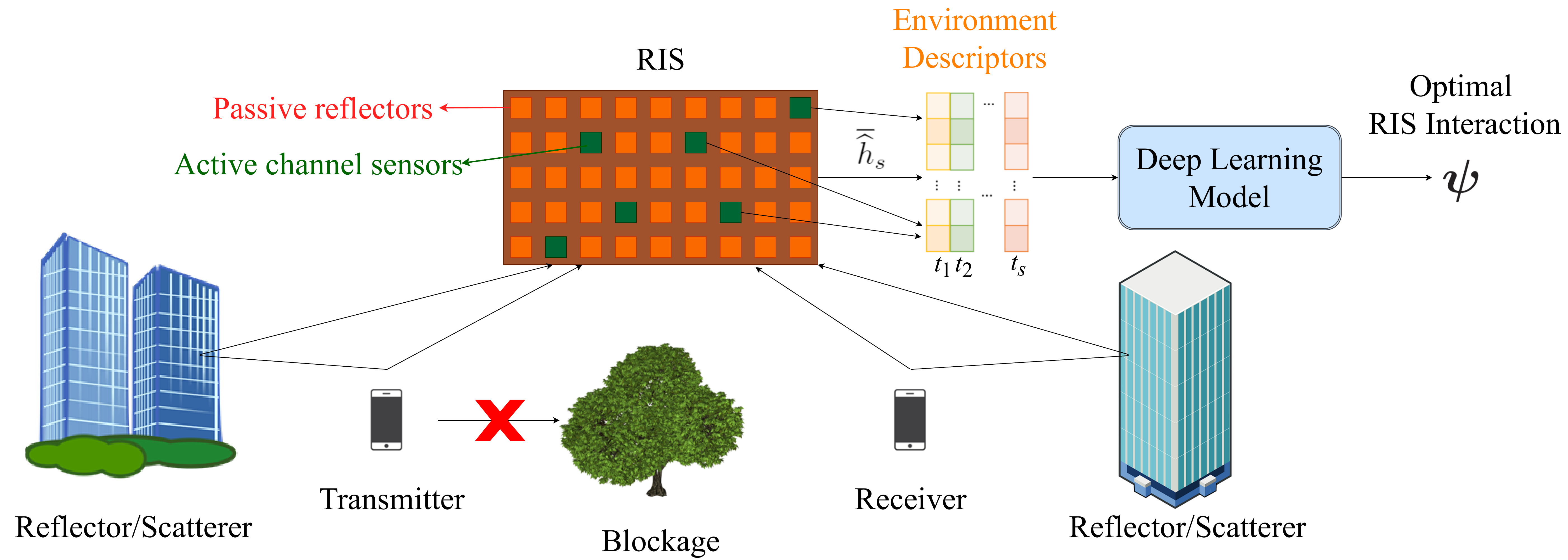}}
\caption{The RIS functionality in a communication environment.}
\label{fig_ML}
\end{figure*}

\section{RIS Phase Interaction Optimization Using Previously Sampled Channels}
\label{Section3}

\subsection{A Motivation}

\par RIS interaction aims at determining the optimal interaction vector $\boldsymbol{\psi}^*$ in the sense of maximizing the receiver achievable information rate. This is achieved by solving
\begin{equation} 
\boldsymbol{\psi}^*=\operatorname*{arg\,max}\limits_{\boldsymbol{\psi} \in \mathcal{P}}\sum^{K}_{k=1}log_2(1+SNR|(\boldsymbol{h}_{T,k}\odot \boldsymbol{h}_{R,k})^T\boldsymbol{\psi}|^2), \label{eq4} 
\end{equation}
\noindent to achieve the optimal rate $R^*$ defined as
\begin{equation} 
R^*=\frac{1}{K}\sum^{K}_{k=1}log_2(1+SNR|(\boldsymbol{h}_{T,k}\odot \boldsymbol{h}_{R,k})^T\boldsymbol{\psi}^*|^2).\label{eq5} 
\end{equation}

\par Finding the optimal interaction vector requires performing an exhaustive search over the codebook provided. In essence, the formulation in (\ref{eq4}) is challenging as one needs to search for an interaction vector for all subcarriers \cite{taha2020deep}. Hence, the costs of obtaining an interaction vector through an exhaustive search are prohibitively large. This is in terms of the costs of training, computation, and power dissipation as detailed in \cite{taha2019enabling}. To this end, it is intuitively sound to think of finding an efficient solution approaching the optimal rate approximately as in (\ref{eq5}). ML-based algorithms are promising to handle complex models. This is especially the case when it is difficult or impossible to have a closed-form mathematical expression of a system \cite{dai2020deep}. Thus, it is intuitive to think of an ML framework for optimizing interaction vectors.

\par
\begin{figure}[b!] 
\setlength\abovecaptionskip{-0.1\baselineskip}
\centering\resizebox{0.92\columnwidth}{!}{
\includegraphics{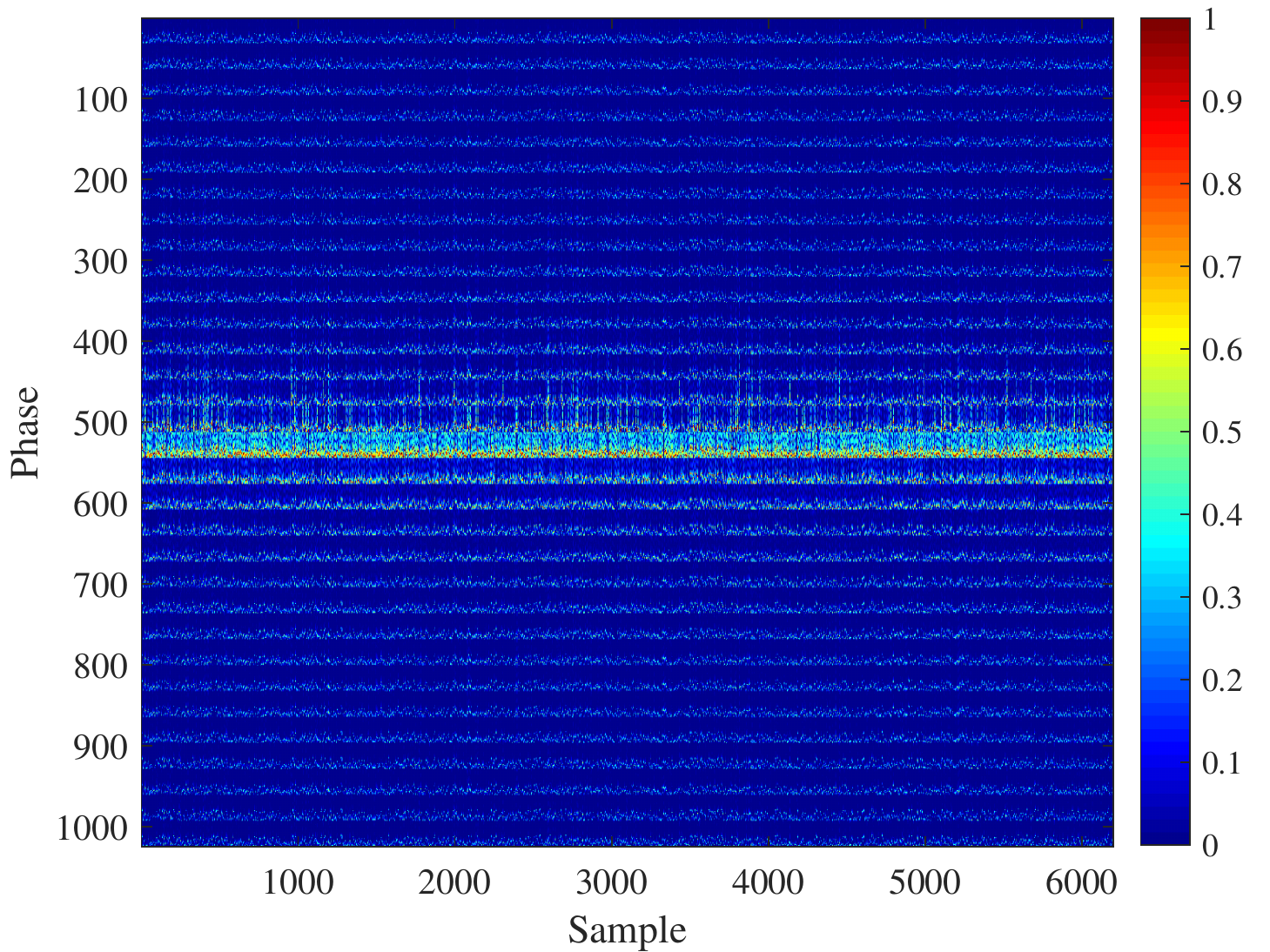}}
\caption{A toy example to show correlation across channel samples.}
\label{motivation}
\end{figure}

 The performance of an ML algorithm depends on the quality of the input and output datasets. In other words, the correlations between the datasets have critical importance for the effectiveness of ML models. In the current literature, phase optimization relies on correlation over the currently sampled channel. On the other hand, the existing inherent correlation over previously sampled channels can be exploited similarly, as a further step. As a preliminary example of this argument, Fig.~\ref{motivation} illustrates the correlations between the current output with the previous output. So, it proves that there is a strong correlation between the current output and the previous outputs. Said conversely, there is a correlation between previously sampled channels with the output to be estimated. Therefore, incorporating the previously estimated sampled channel can improve the performance.

\subsection{The Proposed Algorithm}

\par The proposed algorithm is composed of two stages. First, is a training stage where the DL model is configured and trained. Second is a testing stage whereby optimum phase interaction is carried out using the trained model.

\par In the training stage, the RIS employs an exhaustive search reflection beamforming approach while collecting the dataset for the DL model. After collecting the dataset $t_s-1$, the number of previous channels is concatenated with the current ones, where $t_s$ represents the time step \footnote{For example, if the $t_s$ is set to 3 and there are 1024 optimum phase interactions to be estimated, then each input will have a 3$\times$1024 matrix.}. Once the training dataset is fully acquired, the DL model is trained. The main rationale of these processes is depicted in Fig.~\ref{fig_ML}. Besides, the steps are detailed in Algorithm 1. In this algorithm, the sampled channel vector at every coherence block ($s$) is estimated. Then, the optimal interaction is found by exhaustive beam training (Steps 4 through 6). Afterward, a new data sample for the training dataset is obtained by concatenating the previous channels with the current one (Steps 7 through 9). These processes are repeated until a sufficient dataset is collected. It is noted that we analyze the performance of the proposed algorithm for different sizes of training datasets in the following section. Finally, the DL model is trained according to the input and output datasets.

\par The training stage is followed by the testing stage characterizing the run-time operation of the algorithm. The first step in the testing stage is estimating the sampled channel to be used to find the optimum phase interaction vector. Then, the estimated sampled channel vector and the previously sampled channel vectors are fed to the trained DL model and the reflection beamforming vector is estimated. Algorithm 2 outlines the testing stage of the proposed algorithm.

\subsection{A Note on Computational Complexity}

\par The computational complexity of the proposed algorithm is primarily based on that of the training and testing stages. While the training stage complexity depends on both the exhaustive search and DL model, the testing stage complexity depends only on the DL model. The computational complexity of the exhaustive search is $\bigO(wp)$, where $w$ and $p$ represent the sizes of the codebook \cite{mohammad2006occurrences}. It is noted that this exhaustive search is made for each training sample.

\par A deep MLP model is used in this work with an input layer, three hidden layers, and an output layer, as a DL model. The training computational complexity of this model is $\bigO(nt\times (ij+jk+kl+lm))$ for five layers, where input layer, hidden layers, and output layer represented by $i$, $j$, $k$, $l$, $m$, respectively, and $n$ and $t$ represents number of epochs and training examples, respectively. Furthermore, the per-sample computational complexity of testing is roughly half of the per-sample training computational complexity since the testing stage does not require back-propagation \cite{he2015convolutional}.

\begin{algorithm}[!h]
\caption{The Proposed Algorithm: The Training Stage}
\label{Algorithm1}
\begin{algorithmic}[1]{
\renewcommand{\algorithmicrequire}{\textbf{Input:}}
\renewcommand{\algorithmicensure}{\textbf{Output:}}
\REQUIRE Number of training samples ($S$), two pilots received at the transmitter and receiver for each channel coherence block, codebook ($\mathcal{P}$), and $t_s$.
\ENSURE A trained DL model. \\
\FOR{$s = 1+t_s$ \textbf{to} $S$}
 \STATE{RIS estimates $\overline{\widehat{h}}_t$ and $\overline{\widehat{h}}_r$ using training received pilots.
 }
 \STATE{$\overline{\widehat{h}}_s=\overline{\widehat{h}}_t \odot \overline{\widehat{h}}_r$
 }
\FOR{$n = 1$ \textbf{to} $|\mathcal{P}|$}
 \STATE{RIS reflects beam and receives the feedback $R_n(s)$.
 }
 \ENDFOR 
 \STATE{RIS constructs $r(s) = [R_1(s), R_2(s), \ldots, R_{|\mathcal{P}|}(s)]$. 
 }
 \STATE{Added $t_s$ long historical information as an additional dimension. 
 }
 \STATE{A new data point ($\overline{[\widehat{h}}(s), \overline{[\widehat{h}}(s-1), \ldots, \overline{[\widehat{h}}(s-t_s-1)],r_s \rangle$) is added to the training dataset ($D$).
 }
 \ENDFOR
}
 \STATE{Train the DL model using the generated dataset $D$.}
\end{algorithmic}
\end{algorithm}

\begin{algorithm}[!h]
\caption{The Proposed Algorithm: The Testing Stage}
\label{Algorithm2}
\begin{algorithmic}[1]{
\renewcommand{\algorithmicrequire}{\textbf{Input:}}
\renewcommand{\algorithmicensure}{\textbf{Output:}}
\REQUIRE Two pilots received at the transmitter and receiver, $t_s$, and trained DL model.
\ENSURE The interaction vector. \\
\WHILE{$True$}
 \STATE{RIS estimates $\overline{\widehat{h}}_t$ and $\overline{\widehat{h}}_r$ using testing received pilots.
 } 
 \STATE{$\overline{\widehat{h}}_s=\overline{\widehat{h}}_t \odot \overline{\widehat{h}}_r$ 
 }
 \STATE{Added $t_s$ long historical information as an additional dimension.
 }
 \STATE{The interaction vector is estimated using the trained DL model and sampled channels.
 }
 \ENDWHILE
}
\end{algorithmic}
\end{algorithm}

\section{Simulations and Results}
\label{Section4}
\subsection{Parameter Setting}

\par
\begin{figure}[!t] 
\setlength\abovecaptionskip{-0.1\baselineskip}
\centering\resizebox{0.85\columnwidth}{!}{
\includegraphics{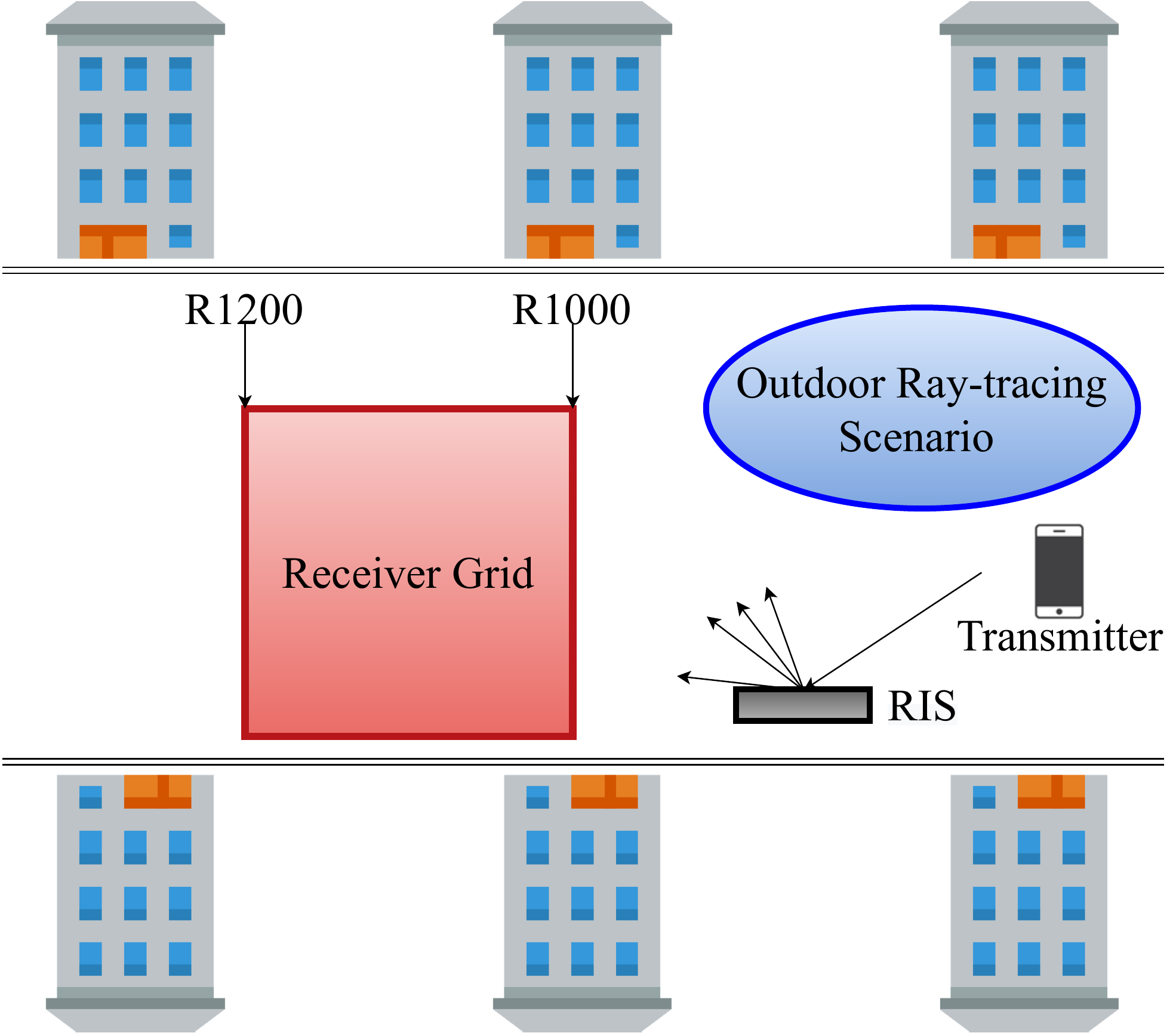}}
\caption{The assumed RIS ray tracing operation \cite{taha2019enabling}.}
\label{fig_1b}
\end{figure}

 We adopt the DeepMIMO dataset\cite{alkhateeb2019deepmimo} to obtain channel realizations. This dataset considers the outdoor ray-tracing scenario ‘O1’. The parameters of this set are listed in Table~\ref{setup_table}. It is noted that the values of ($M_x$, $M_y$, $M_z$), $P_T$, and $L$ are changed for some simulations and these parameter changes are specified where the simulations are presented. According to this setting, the transmitter is assumed to be fixed, whereas the receiver's location is randomly distributed in the x-y plane as illustrated in Fig.~\ref{fig_1b}. It is noted that this scenario is generated according to Remcom Wireless InSite \cite{ray_tracing}, and is publicly available on the DeepMIMO dataset \cite{alkhateeb2019deepmimo}. The RIS is chosen to be BS 3, as further described in \cite{taha2019enabling}. An optimum $t_s$ is empirically set to 3 based on the performance and generalization capability of the proposed algorithm.

\begin{figure*}[!t]
\centering
\centering\resizebox{1.85\columnwidth}{!}{
\includegraphics{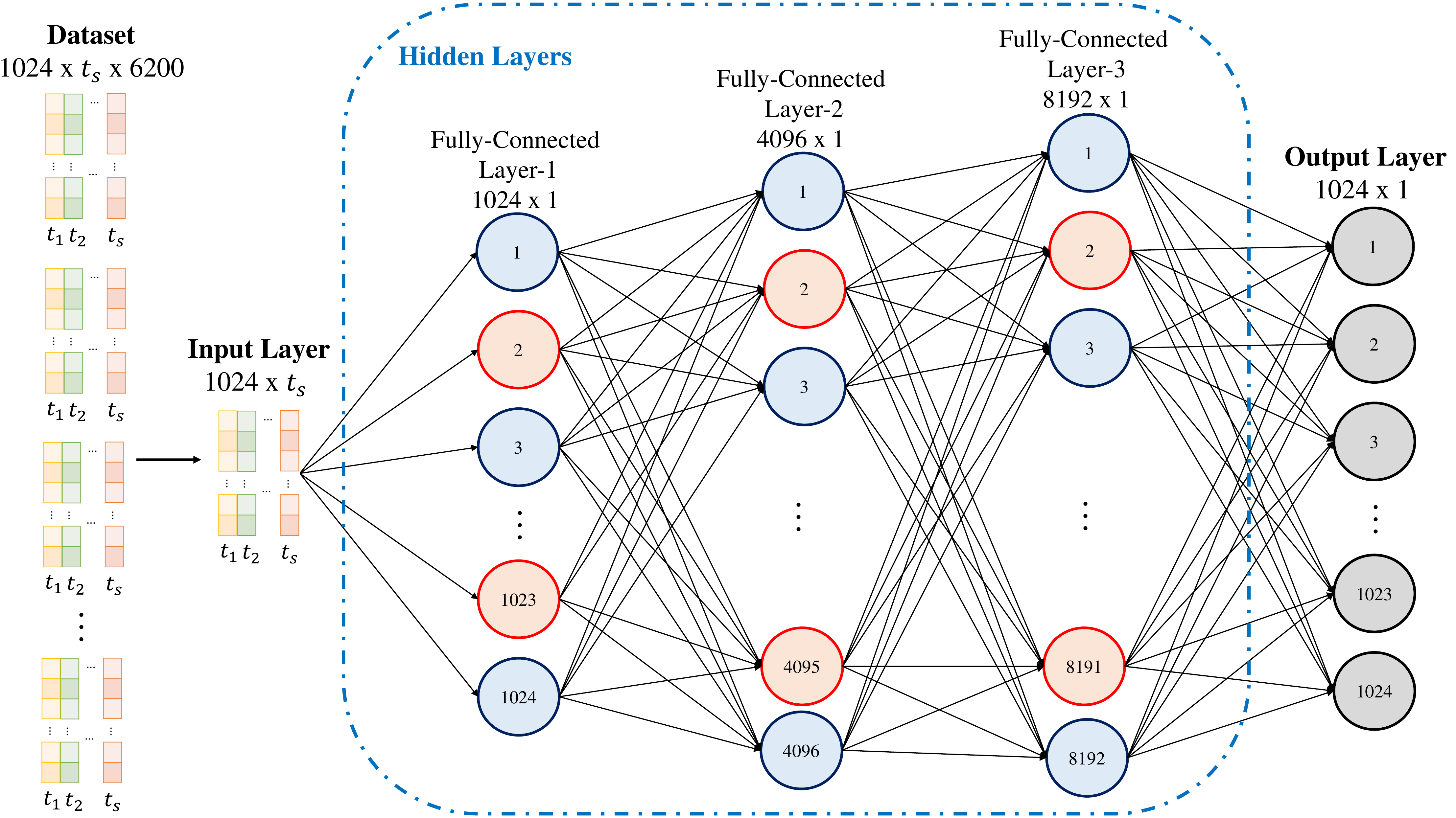}}
\linespread{1}
\caption{Incorporating previous channel information in RIS phase optimization.}
\label{mlagorithm}
\end{figure*}

\begin{table}[!htb]
\centering
\caption{Parameters of the DeepMIMO dataset.}
\begin{tabular}{|c|c|}
\hline
\textbf{Property} & \textbf{Value} \\
\hline\hline
Frequency band & 28 GHz \\ \hline
System bandwidth & 100 MHZ \\ \hline
Antenna spacing & 0.5$\lambda$ \\ \hline
Number of BS Antennas & ($M_x$, $M_y$, $M_z$) = (1, 32, 32) \\ \hline
Active transmitter & Row R850 column 90 \\ \hline
Active receivers & From row R1000 to row R1200 \\ \hline
Active BSs & 3 \\ \hline
Number of paths & 1 \\ \hline 
Number of OFDM subcarrier & 512 \\ \hline
OFDM limit & 64 \\ \hline
OFDM sampling factor & 1 \\ \hline
\end{tabular}
\label{setup_table}
\end{table}

\begin{figure}[b!] 
\setlength\abovecaptionskip{-0.1\baselineskip}
\centering\resizebox{0.92\columnwidth}{!}{
\includegraphics{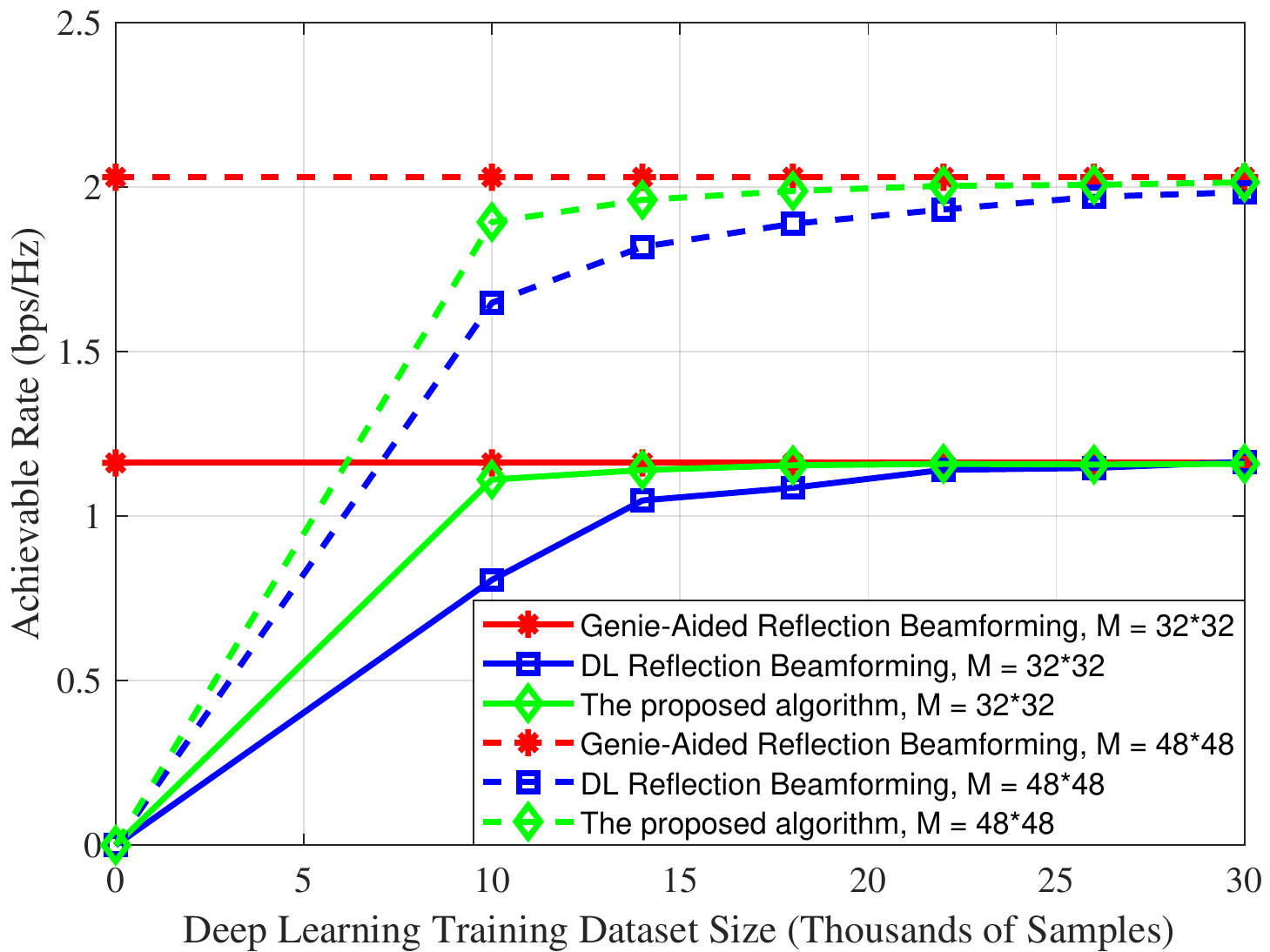}}
\caption{The achievable rate comparison between the algorithm in \cite{taha2019enabling} and the proposed algorithm when $M=32\times32$ and $M=48 \times 48$.}
\label{DifferentDatasets1}
\end{figure}

\subsection{Hyperparameters of Deep Learning Model}

\par The DL architecture for RIS interaction optimization is implemented by the MATLAB R2019a DL Toolbox. All of the hyperparameters are tuned empirically by considering the performance and generalization capability of the proposed algorithm. In the MLP model, an input layer, three hidden layers (fully-connected layers), and an output layer are used. More specifically, 1024 units are used in the input layer. 1024, 4096, and 8192 hidden units are used in the first, second, and third hidden layers, respectively, and 1024 units are used in the output layer to find optimum phase interactions. In all of these layers, the rectified linear unit is used as an activation function, and after all hidden layers, a dropout is used with a 0.5 factor to prevent overfitting. The regression layer is used to estimate numerical optimal RIS interaction in the output layer. It is noted that only the first $K=64$ subcarriers are fed to the DL model as in \cite{taha2019enabling}.

\par The maximum number of epochs is selected to be 49, the batch size is 500, and $\ell_{2}$ regularization is used with a value of 0.0001. Besides, the initial learning rate is selected as 0.1 and dropped in every 8 iterations with a 0.5 factor. The dataset is split into two sets; training and testing. In the testing stage, always 6200 samples are used, and the number of samples used in the training is varied in all simulations. The structure of the DL model used is illustrated in Fig.~\ref{mlagorithm}.

\begin{figure}[b!] 
\setlength\abovecaptionskip{-0.1\baselineskip}
\centering\resizebox{0.92\columnwidth}{!}{
\includegraphics{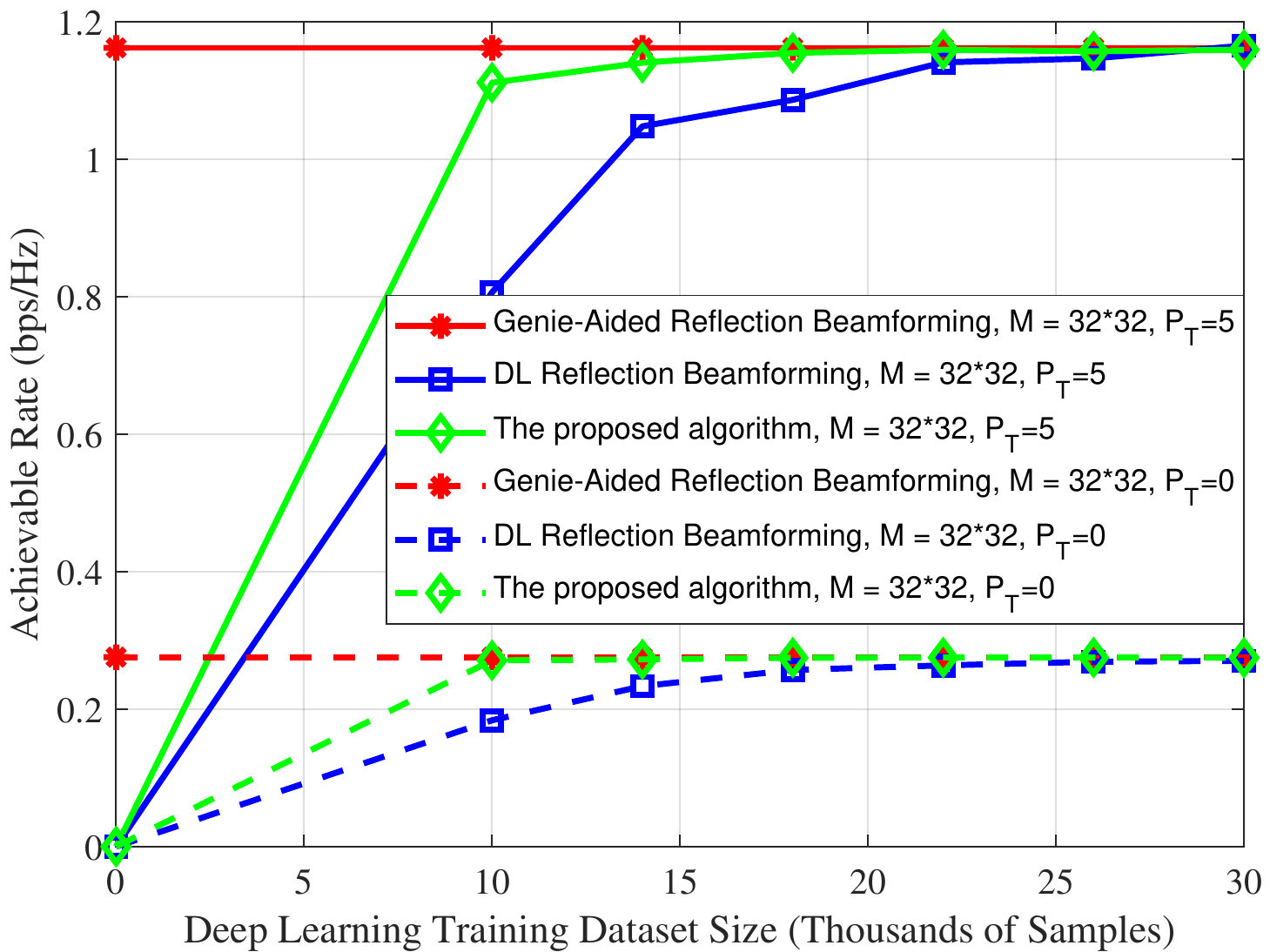}}
\caption{The achievable rate comparison between the algorithm in \cite{taha2019enabling} and the proposed algorithm when $M=32 \times 32$, $P_T=5$ and $P_T=0$.}
\label{DifferentDatasets2}
\end{figure}

\subsection{Performance Evaluations and Discussions}

\par The optimal RIS interaction selection performance is evaluated in terms of the achievable rate for various scenarios. In this regard, we compare the algorithms in \cite{taha2019enabling} with the proposed algorithm.

\par
\begin{figure}[t!] 
\setlength\abovecaptionskip{-0.1\baselineskip}
\centering\resizebox{0.92\columnwidth}{!}{
\includegraphics{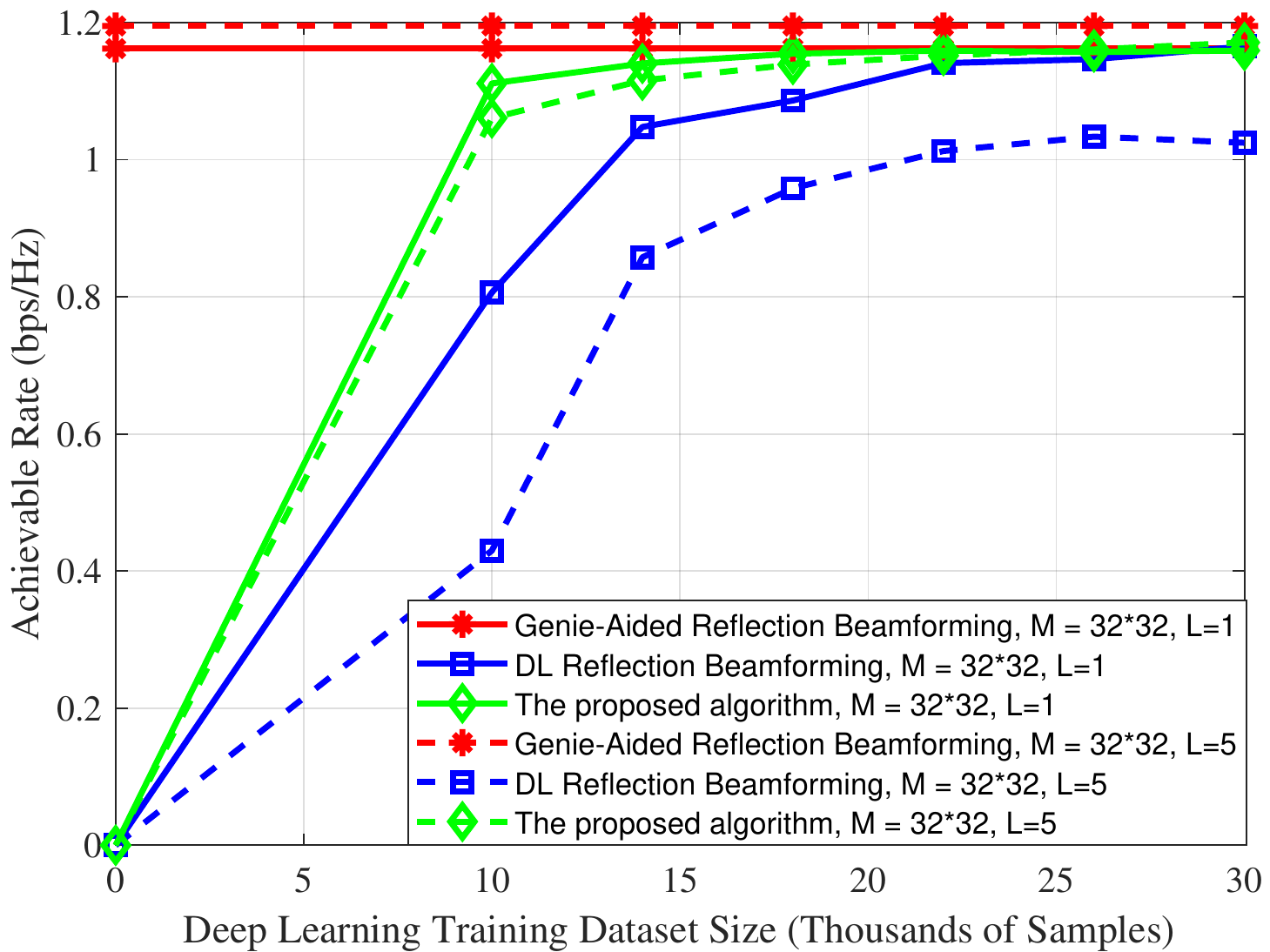}}
\caption{The achievable rate comparison between the algorithm in \cite{taha2019enabling} and the proposed algorithm when $M=32 \times 32$, $L=1$ and $L=5$.}
\label{DifferentDatasets3}
\end{figure}

 Fig.~\ref{DifferentDatasets1} represents the achievable rate performance of the proposed algorithm for optimum phase interaction when the RIS operates with either a 48$\times$48 or a 32$\times$32 uniform planar array (UPA). Fig.~\ref{DifferentDatasets1} shows that the proposed algorithm is superior to the algorithm in \cite{taha2019enabling} when the UPA is 48$\times$48 and 32$\times$32. Also, the performance of the proposed algorithm is evaluated for different $P_T$ and $L$ values in Figs.~\ref{DifferentDatasets2} and \ref{DifferentDatasets3}, respectively. These figures also reveal the effectiveness of the proposed algorithm. This is especially the case when the amount of training data is low. 

\par
\begin{figure}[htb!] 
\setlength\abovecaptionskip{-0.1\baselineskip}
\centering\resizebox{0.89\columnwidth}{!}{
\includegraphics{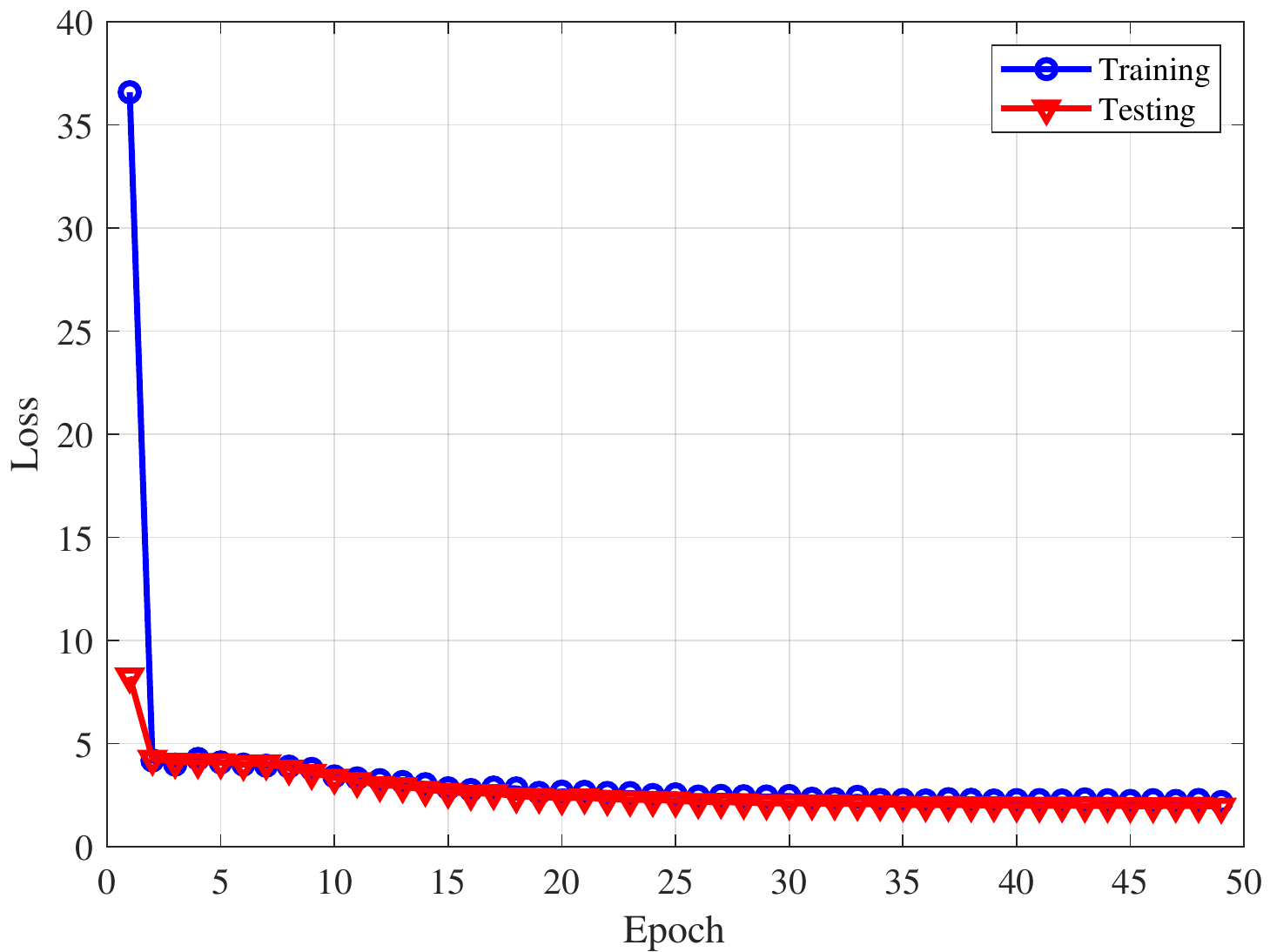}}
\caption{The loss graph of the proposed algorithm when $M=32\times32$, $L=1$, and $P_T=5$.}
\label{loss}
\end{figure}

 As the proposed phase interaction algorithm is based on DL, it is important to verify that the established model does not memorize the inputs during the training stage. To investigate this, the testing and training losses versus epochs for optimum phase interaction are plotted in Fig.~\ref{loss} when ($M_x$, $M_y$, $M_z$)=(1, 32, 32), $L$=1, $P_T$=5, and the training and testing dataset sizes are 20000 and 6200, respectively. As can be seen in the figure, the loss of the testing set converges to that of the training set. This verifies the nonexistence of overfitting, so validating the generalizability of the proposed algorithm. It is noted that the loss graph is provided solely for one scenario. Similar behavior is observed for the graphs of other scenarios.

\section{Conclusions}
\label{Section5}

\par In this paper, RIS-assisted wireless communication systems are considered and an algorithm for phase optimization is proposed. This is based on exploiting an existing correlation between the current and previously estimated channels. A DL model is designed to exploit this correlation to increase the achievable rate. The increase of the achievable rate was revealed by extensive simulations conducted over a ray-tracing dataset.

\section*{Acknowledgement}
\par The work of H. Arslan was supported by the Scientific and Technological Research Council of Turkey (TUBITAK) under Grant 119E433.

\end{document}